\documentclass[twocolumn,pre,floatfix]{revtex4}
\usepackage{psfrag,epsfig,amsfonts,amssymb,amsmath,wasysym,bm}
\usepackage{dcolumn}

\usepackage[normalem]{ulem}
\usepackage{color}

\renewcommand{\i}{{\rm{i}}}
\newcommand{\hh}{\hbar}
\newcommand{\rhobar}{\overline{\rho}}

\newcommand{\lu}{\left[}
\newcommand{\ru}{\right]_{\Pi}}
\newcommand{\IE}{I_E}

\newcommand{\lsim}{\mathrel{\hbox{\rlap{\lower.55ex \hbox{$\sim$}} \kern-.3em \raise.4ex \hbox{$<$}}}}
\newcommand{\gsim}{\mathrel{\hbox{\rlap{\lower.55ex \hbox{$\sim$}} \kern-.3em \raise.4ex \hbox{$>$}}}}

\newcommand{\kB}{k_{\rm B}}

\newcommand{\DD}{D}

\newcommand{\propa}{{\cal U}_t}

\newcommand{\tr}{\mbox{Tr}}
\newcommand{\da}{\Delta_{\! A}}

\newcommand{\Jmax}{J_{\rm{max}}}
\newcommand{\Teff}{T_{\rm{eff}}}
\newcommand{\HD}{\tilde D(t)}

\providecommand{\norm}[1]{\|#1\|}

\newcommand{\CC}{{\mathbb C}}
\newcommand{\NN}{{\mathbb N}}

\begin{document}

\title{Typical Relaxation of Isolated 
Many-Body Systems Which Do Not Thermalize}

\author{Ben N. Balz}
\author{Peter Reimann}
\affiliation{Fakult\"at f\"ur Physik, 
Universit\"at Bielefeld, 
33615 Bielefeld, Germany}

\begin{abstract}
We consider isolated many-body quantum systems which do not 
thermalize, i.e., expectation values approach an (approximately) 
steady longtime limit which disagrees with the microcanonical 
prediction of equilibrium statistical mechanics.
A general analytical theory is worked out for the
typical temporal relaxation behavior in such cases.
The main prerequisites are initial conditions 
which appreciably populate many energy 
levels and do not give rise to 
significant spatial inhomogeneities on macroscopic scales.
The theory explains very well the
experimental and numerical findings
in a trapped-ion quantum simulator
exhibiting many-body localization, 
in ultracold atomic gases, 
and in integrable hard-core 
boson and XXZ models.
\end{abstract}

\maketitle

The long-standing task to explain macroscopic 
equilibration phenomena in terms of the 
underlying microscopic quantum dynamics 
is presently regaining considerable attention
\cite{rev1,nan15,rev2}.
Since open systems are beyond the realm
of standard quantum mechanics, the common 
starting point is an isolated many-body 
system, possibly incorporating the 
environment of the subsystem of 
actual interest.
The question whether and how such a 
system or subsystem approaches some thermal 
or nonthermal equilibrium state after a 
sufficiently long time has been at the 
focus of numerous
analytical
\cite{ana,eth1,pal15},
numerical 
\cite{rig08,eth2,rig09,num,pal10,gog11},
and experimental 
\cite{gri12,kuh13,rau16,smi16,mblexp,exp}
studies.
Despite the reversible and everlasting
motion of the microscopic degrees of freedom,
it could be shown in Refs. \cite{relax,sho12}
under increasingly weak assumptions
about the system Hamiltonian, the initial
condition, and the considered observable
that expectation values must
remain extremely close to a constant value
for the vast majority of all sufficiently
late times (the exceptional times
include initial transients and quantum 
revivals).

The natural next question is whether
the system thermalizes, that is, whether
the longtime behavior is well approximated 
by the pertinent microcanonical expectation 
value from equilibrium statistical 
mechanics.
A first prominent criterion for
thermalization is 
%the validity of 
the so-called eigenstate 
thermalization hypothesis (ETH),
postulating
%that every energy eigenstate is thermal,
%i.e., its expectation values are close 
that every energy eigenstate yields 
expectation values close to the corresponding 
microcanonical values  
\cite{eth1,pal15,rig08,eth2}.
In other words, a violation of
ETH is commonly considered 
%as one key signature of 
%to imply 
an indicator of nonthermalization 
\cite{rev1,pal15,rig08,eth2,rig09}.
A related but different such indicator 
%of non-thermalization 
is the existence of additional 
conserved quantities (besides the 
system Hamiltonian) which can be
written as sums of local operators, 
and which play a particularly
prominent role for so-called
integrable systems
\cite{rev1}.
Numerically, it has been found 
that such systems usually violate 
the ETH and do not thermalize 
\cite{rig08,eth2,rig09}. 
Instead, the longtime behavior
is well captured by a so-called 
generalized Gibbs ensemble (GGE),
which is obtained by the standard
working recipe to maximize the von 
Neumann entropy under the constraints 
that the expectation values of 
the conserved quantities 
%$H$ and $C_k$ 
must be correctly reproduced \cite{rig08}.
Yet another common distinction 
between integrable and nonintegrable 
systems is the statistics of the gaps 
between neighboring energy levels $E_n$
\cite{rev1}.
Further prominent examples which do
not thermalize are systems exhibiting 
many-body localization (MBL) 
\cite{nan15,pal10,gog11,smi16}.
Compared to integrable systems, 
they are structurally more robust 
against small changes of the model
Hamiltonian,
%\cite{nan15},
but they otherwise seem to be quite 
similar, e.g., regarding
energy level statistics,
conserved quantities,
ETH violation,
and the GGE 
\cite{rev1,nan15,mblnum}.

The objective of our Letter is 
%to develop an analytical theory 
%Here, our main objective will be
a quantitative analytical description 
of the temporal relaxation in the 
absence of thermalization.
%The main objective of our present 
%Letter is to derive general analytical 
%results regarding the detailed temporal 
%relaxation of systems which exhibit 
%equilibration but do not thermalize.
Our approach is thus complementary to 
%a variety of related numerical case studies
the numerical case studies, e.g., 
in Refs. \cite{rig09,gog11,mblnum,wu16}.
Related analytical investigations
are also quite numerous
%Furthermore, there also exist
%very impressive analytical 
%investigations on similar questions 
\cite{sho12,times,rei16}.
Yet, for each of them, a closer look 
at the considered systems and the 
obtained results reveals quite 
significant differences from our 
approach.
For instance, some of them 
concern only thermalizing systems,
%[],
%implicitly or explicitly 
%exclude systems which do not thermalize,
others focus on special observables,
%[]
or on deriving
%but far from tight
upper and lower 
bounds for the temporal relaxation, 
%[]
etc.
Particularly little is known about 
equilibration time scales in isolated 
systems which do not thermalize.
Likewise, pertinent experimental works 
are still rather scarce 
\cite{gri12,kuh13,rau16,smi16,mblexp}.
A comparison of our theory with 
exemplary numerical and experimental
results is provided later.

Going {\em in medias res}, let us consider 
a Hamiltonian $H$ with eigenvalues 
$E_n$ and eigenvectors $|n\rangle$
and an arbitrary initial state $\rho(0)$
(pure or mixed and, in general, far 
from equilibrium).
According to textbook quantum 
mechanics, its temporal evolution is
$\rho(t)=\propa\rho(0)\propa^\dagger$
with $\propa:=e^{-\i Ht/\hh}$.
Hence, the expectation value
$\langle A\rangle_{\!\rho}:= \tr\{\rho A\}$
of an arbitrary observable $A$ follows as
\begin{eqnarray}
\langle A\rangle_{\!\rho(t)}
= \sum_{m,n}
\rho_{mn}(0) A_{nm}
\, e^{\i (E_n-E_m)t/\hh} 
\ ,
\label{10}
\end{eqnarray}
where $A_{mn}:=\langle m |A| n \rangle$,
$\rho_{mn}(t):=\langle m|\rho(t)|n\rangle$
and where,
depending on the specific model under 
consideration, $m$ and $n$ run from $1$ 
to infinity or to some finite upper limit.
Averaging Eq. (\ref{10}) over all 
%times 
$t\geq 0$
yields the result $\langle A\rangle_{\!\rhobar}$,
where the 
%so-called 
diagonal ensemble $\rhobar$ is defined via 
$\rhobar_{mn}:=\delta_{mn} \rho_{nn}(0)$
\footnote{If $H$ exhibits degeneracies,
we tacitly choose the eigenvectors
$|n\rangle$ so that $\rho_{mn}(0)$ 
is diagonal within every eigenspace.}.
Hence, if the system equilibrates 
at all, (\ref{10}) must stay extremely 
close to $\langle A\rangle_{\!\rhobar}$
for practically all sufficiently large 
%times 
$t$ (see above).

As usual, we focus on systems with
a macroscopically well-defined energy; i.e.,
%In doing so, it is tacitly assumed that
%the system exhibits a macroscopically 
%well-defined energy, i.e., 
all energy levels $E_n$ with non-negligible 
populations $\rho_{nn}(0)$ 
%notably contribute to the sum 
%$\sum_n \rho_{nn}(0)=1$ 
must be contained in an
%macroscopically very small 
%(but microscopically large) energy 
interval $\IE:=[E-\epsilon,\,E]$
of macroscopically small 
(but microscopically large) 
width $\epsilon$.
Furthermore, we adopt the idealization 
that the probability $\rho_{nn}(0)$
to observe an energy $E_n$ outside $\IE$ 
can be approximated as strictly zero.
The number of energies $E_n$ contained 
in $\IE$ is denoted by $D$ and, without loss
of generality, we assume that 
$n\in\{1,...,D\}$ for all those $E_n$.
The Cauchy-Schwarz inequality 
$|\rho_{mn}|^2\leq \rho_{mm}\rho_{nn}$
then implies that only $m,\,n\leq D$
actually matter in Eq. (\ref{10})
and in all that follows.
Specifically, the effectively 
relevant Hamiltonian is
$H_1:=\sum_{n=1}^D E_n\,|n\rangle\langle n|$.

Denoting by $\pi$ any permutation of $\{1,...,D\}$,
we define
\begin{eqnarray}
H_\pi:=
%\sum_{n=1}^D E_{\pi(n)}\,|n\rangle\langle n|
%=
%\sum_{n=1}^D E_{n}\,|\pi^{-1}(n)\rangle\langle \pi^{-1}(n)|
\sum_{n=1}^D E_{n}\,|\pi(n)\rangle\langle \pi(n)|
=
\sum_{n=1}^D E_{\pi^{-1}(n)}\,| n\rangle\langle n|
\ .
\label{20}
\end{eqnarray}
%%with $n_\pi:=\pi^{-1}(n)$.
%The last identity follows from
%%$|n\rangle=|\pi^{-1}(\pi(n))\rangle$ 
%$n=\pi^{-1}(\pi(n))$ 
%and adopting $\pi(n)$ as new summation index.
Hence, 
$H_\pi$ is obtained by permuting either
the eigenvalues or the eigenstates
of the original Hamiltonian $H_1$.

In general, every $H_\pi$ entails a 
different evolution of $\rho(t)$.
Accordingly, in Eq. (\ref{10}) either the 
energies or the matrix elements must be 
permuted analogously as in (\ref{20}).
On the other hand, one readily 
%verifies the following important 
%invariances under arbitrary
sees that the following 
important quantities and properties 
are invariant under arbitrary
%characteristics remain unchanged for all
permutations $\pi$:
%basic properties are 
%common to all 
%Hamiltonians 
%$H_\pi$:
(i) the energy spectrum, and hence the 
level statistics; 
(ii) the violation or 
nonviolation of the ETH; 
(iii) 
the conserved quantities 
\footnote{Any conserved quantity 
$C$ satisfies $[C,H_1]=0$ and thus
implies a common eigenbasis of $C$ and $H_1$.
For a nondegenerate $H_1$, this basis 
must be $\{|n\rangle\}_{n=1}^D$,
and with Eq. (\ref{20}), it follows that
$[C,H_\pi]=0$.
If $H_1$ exhibits degeneracies,
$H_\pi$ still possesses just as many
conserved quantities as $H_1$,
but, in general, they are no 
longer identical.}; 
(iv) 
the initial expectation value
$\langle A\rangle_{\!\rho(0)}$.
%The diagonal ensemble $\rhobar$.
%and all the concomittant equilibration 
%properties (see below (\ref{10})).
%For any given $H_\pi$ 
%Hence, 
(v) 
for the vast majority of all
sufficiently large $t$,
the expectation value 
$\langle A\rangle_{\!\rho(t)}$
stays extremely close to
$\langle A\rangle_{\!\rhobar}$, 
with the same diagonal ensemble
$\rhobar$ for all 
$H_{\pi}$,
%permutations $\pi$,
%as for $H_1$,
and likewise for the GGE.

The main result of our Letter concerns 
the $\pi$ and $t$ dependent relaxation 
of $\langle A\rangle_{\!\rho(t)}$ 
and reads
\begin{eqnarray}
\langle A\rangle_{\!\rho(t)} 
& = & 
\langle A\rangle_{\!\rhobar} 
+
F(t)\, \left\{ \langle A\rangle_{\!\rho(0)} 
- \langle A\rangle_{\!\rhobar}\right\}
+ \xi_\pi(t) \ ,
\label{30}
%\end{eqnarray}
%where $F(t)$ is given by
%\begin{eqnarray}
\\
F(t) & := & \left(\DD \,|\phi(t)|^2-1\right)/(D-1) \ ,
\label{40}
\\
\phi(t) & := & 
\DD^{-1}\mbox{$\sum_{n=1}^{\DD}$}e^{\i E_n t/\hh} 
\ .
\label{50}
\end{eqnarray}
%The entire $\pi$-dependence of (\ref{30})
%is thus encapsulated in 
The only $\pi$ dependent term on the right-hand 
side of Eq. (\ref{30}) is
$\xi_\pi(t)$ and 
satisfies, for $D\geq 6$, the following key properties:
\begin{eqnarray}
\lu \xi_\pi(t) \ru & = & 0 \ ,
%\label{60}
%\\
%\lu \xi^2_\pi(t) \ru & \leq & 
\ \lu \xi^2_\pi(t) \ru \leq
(6\da)^2 \max_n\rho_{nn}(0) 
%\tr\rhobar^2
\ ,
\label{60}
\end{eqnarray}
where $\Pi$ denotes the set of all
permutations of $\{1,...,D\}$
and $[...]_\Pi$ 
the average over all $\pi\in\Pi$.
Furthermore, $\da$ is the measurement 
range of the observable $A$, i.e.,
the difference between its 
largest and smallest eigenvalues.

Equations (\ref{30})-(\ref{60}) are exact 
analytic results
when $D\geq 6$ and for arbitrary $H$, $A$, and $\rho(0)$
with $\rho_{nn}(0)=0$ for $n>D$.
%The detailed mathematical derivation 
%of these results 
Their detailed mathematical derivation
is quite tedious 
and provides very little physical insight;  
hence, it has been postponed 
to the Supplemental Material.
%\cite{sup}.

Since a typical many-body system 
exhibits an extremely dense energy 
spectrum (exponential in the degrees 
of freedom), 
it is practically impossible (e.g., in an 
experiment) to notably populate only a few 
energy levels; hence,
$\max_n \rho_{nn}(0)$
must be unimaginably small
\cite{relax}.
Observing that 
%$1/\DD\leq \tr\rhobar^2 \leq \max_n\rho_{nn}(0)$
$1/\DD\leq \max_n\rho_{nn}(0)$
implies that 
$D\gg 1$ in Eqs. (\ref{40}) and (\ref{50}),
that the number $D!$ of permutations 
$\pi\in\Pi$ is gigantic,
and that $\lu \xi^2_\pi(t) \ru$ 
in Eq. (\ref{60}) is exceedingly small.
As a consequence, $\xi_\pi(t)$ itself must be 
very small for the vast majority of 
all $\pi\in\Pi$; i.e., we can safely
approximate Eq. (\ref{30}) by
\begin{eqnarray}
\langle A\rangle_{\!\rho(t)} 
& = & 
\langle A\rangle_{\!\rhobar} 
+
F(t)\, \left\{ \langle A\rangle_{\!\rho(0)} 
- \langle A\rangle_{\!\rhobar}\right\}
\ .
\label{70}
\end{eqnarray}
Specifically, this approximation also
applies to the ``true'' system
$H_1$, unless there are special reasons 
why its temporal relaxation should 
notably differ from that of practically
all other $H_{\pi}$.

A first very strong argument why the true system 
may be expected to exhibit the typical relaxation 
behavior (\ref{70}) is the abovementioned 
invariances (i)-(v) under arbitrary 
permutations $\pi$.
%properties (i)-(v), which are common to all 
%Hamiltonians $H_{\pi}$ from (\ref{20}).
%When considering all those Hamiltonians as 
In fact, when considering the corresponding 
Hamiltonians $H_\pi$ as a matrix ensemble, 
our situation is essentially just a 
particular instance of random matrix theory
\cite{bro81},
whose predictions are well known to be surprisingly 
successful in many cases, provided that the ensemble 
%correctly accounts for 
preserves a few very basic properties 
of the true system of actual interest
\cite{bro81}
(e.g. symmetries, or the 
%properties 
invariances (i)-(v) in our case).
%Below, we will further corroborate
%this working hypothesis by comparison 
%with experimental and numerical results.
%We further note that
%though most Hamiltonians of the ensemble
%are in many other respects quite different 
%from the ``true'' system of actual interest.
%or even physically meaningless.

On the other hand, usual model 
Hamiltonians $H_1$ only involve 
%few-body interactions of short range 
%and thus can be written as sums of 
%local operators 
short-range interactions 
(or local operators)
\cite{rev1,nan15},
while most other $H_\pi$ do not 
preserve this ``local structure''.
Spatial inhomogeneities 
of particle numbers, energy
etc. are thus expected to 
be balanced out increasingly 
slowly over increasing distances
when $H_1$ governs the dynamics,
but not for most other $H_\pi$.
Note that instead of permuting the energy 
eigenvectors $|n\rangle$ in Eq. (\ref{10}) 
according to Eq. (\ref{20}), one could 
replace $\rho(0)$ by
$\rho_\pi(0):=U_\pi^\dagger\rho(0)U_\pi$,
where the unitary $U_\pi$ is defined
via $U_\pi|n\rangle=|\pi(n)\rangle$
(and likewise for $A$, while $H_1$ 
is now kept fixed).
Once again, even when $\rho(0)=\rho_1(0)$
exhibits spatial inhomogeneities,
one expects that most other $\rho_\pi(0)$ 
will appear (nearly) homogeneous;
hence, the local structure of $H_1$
yields an untypically slow
relaxation of $\rho(0)$ 
(compared to most other 
$\rho_\pi(0)$).
In either case, it follows that 
our prediction (\ref{70}) 
must be restricted to initial 
conditions without any significant 
spatial inhomogeneities on 
macroscopic scales.

A typicality result similar to 
Eqs. (\ref{30})-(\ref{60}) was 
obtained by formally quite
dissimilar methods in 
Ref. \cite{rei16}.
Conceptually, the essential
difference is that arbitrary 
unitary (Haar distributed) 
basis transformations rather than 
just eigenvector permutations 
in Eq. (\ref{20}) were admitted in 
Ref. \cite{rei16},
resulting in the appearance 
of the microcanonical instead 
of the diagonal ensemble 
on the right-hand side of 
Eq. (\ref{30}).
In contrast to our work,
the approach from Ref. \cite{rei16}
is thus restricted to systems 
which do thermalize. 
%The main step forward of our present
%work is that it applies to systems 
%which do {\em not} thermalize.
%\footnote{Incidentally, it also applies to 
%systems which do thermalize, provided 
%the diagonal and microcanonical 
%ensembles yield identical expectation 
%values, e.g. due to the validity of ETH}.
The main reason is that the permutations
are a tiny subset (of measure zero) 
of all unitary basis transformations
and thus may preserve additional
key features of the 
true Hamiltonian $H_1$.
For example, permutations preserve 
each of the abovementioned properties (i)-(v),
but general unitaries preserve only  (i) and (iv).
In return, the smallness of 
$\max_n\rho_{nn}(0)$ on the right-hand 
side of Eq. (\ref{60}) is no longer 
required when admitting arbitrary
unitaries \cite{rei16}.
In passing, we note that conditions
similar to or even identical 
to $\max_n\rho_{nn}(0)\ll 1$
already arise in the general equilibration 
results from Refs. \cite{relax,sho12}.

Turning to the function $F(t)$ in 
Eqs. (\ref{40}) and (\ref{50}), one readily 
sees that $F(0)=1$ and $1\geq F(t)>-1/D$ 
for all $t$.
Moreover, the following properties 
%of $F(t)$ 
were derived previously in Ref. \cite{rei16}:
(i) $F(t)$ remains negligibly small
for the vast majority of all 
sufficiently large $t$, provided the 
maximal energy degeneracy 
is much smaller than $D$.
(ii) Denoting by $\Omega(E)$ the number of
energies $E_n$ below $E$, by $\kB$ and
$S(E)=\kB\ln \Omega(E)$ 
Boltzmann's constant and entropy,
respectively, and by $T:=1/S'(E)$
the corresponding formal temperature,
one can often approximate the sum 
in Eq. (\ref{50}) by an integral over
a suitably smoothened level density,
yielding the approximation
\begin{eqnarray}
F(t)=1/[1+(t\,\kB T/\hbar)^2] \ .
\label{80}
\end{eqnarray}
%with $\tau:=\hbar/\kB T$.
Note that $T$ and $S(E)$ could be identified 
with the usual temperature and entropy 
%if the system were in thermal equilibrium
for a thermalized system,
%(at energy $E$), 
but they have no immediate physical meaning
for nonthermalizing systems.
%these are purely formal quantities 
%without any clear physical meaning.

Besides integrability and MBL, yet another 
(quite trivial) reason for nonthermalization 
may be that the non-negligible level populations 
$\rho_{nn}(0)$ are not confined to a
macroscopically small energy interval  
(see above Eq. (\ref{20})).
Incidentally, this case can also be readily
included in our present theory, namely, 
by choosing $D$ and the labels $n$ so 
that $n\in\{1,...,\DD\}$ 
if and only if $\rho_{nn}(0)$ is 
non-negligible. 
As a consequence, Eq. (\ref{80}) is, in general,
no longer valid, while all other 
findings remain essentially 
unchanged.

\begin{figure}
\epsfxsize=1.0\columnwidth
\epsfbox{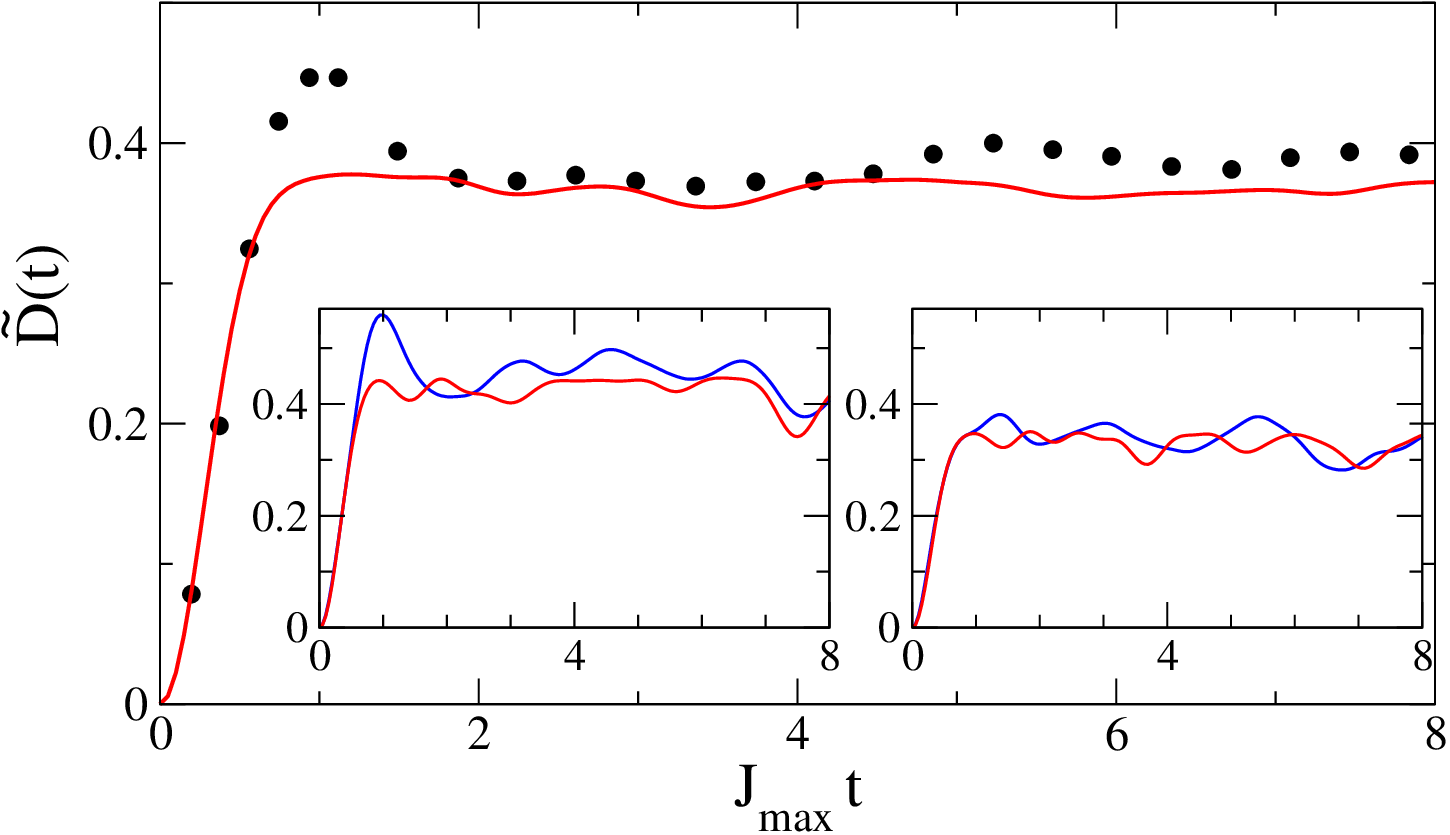}
\caption{\label{fig1}
Symbols: The experimentally measured 
Hamming distance $\HD$ from Fig. 3(a) 
of Ref. \cite{smi16} for $W=4\Jmax$, 
averaged over 30 realizations of the 
disorder in Eq. (\ref{90}).
Line: Corresponding theory from Eq. (\ref{70}).
Insets: Theory (red curves) and
numerical solutions (blue curves) 
for two representative realizations 
of the disorder in Eq. (\ref{90}).
%For further details see main text.
}
\end{figure}

As a first example, we consider the experiment by
Smith et al. \cite{smi16} with $N=10$ ions in a 
linear Paul trap, 
%simulating in very good approximation
%realizing 
%mimicking
emulating
the disordered Ising Hamiltonian
\begin{eqnarray}
H=\sum_{i<j}J_{ij}\sigma_i^x\sigma_j^x
+\frac{B}{2}\sum_i\sigma_i^z
+\sum_i\frac{D_i}{2}\sigma_i^z
%+\sum_i\frac{B+B_i}{2}\sigma_i^z
\label{90}
\end{eqnarray}
with $i,j=1,...,N$, the Pauli matrices
$\sigma_i^{x,z}$,
%acting on spin $i$,
the couplings $J_{ij}=\Jmax/|i-j|^{1.13}$,
the homogeneous field $B=4\Jmax$,
the uniformly distributed 
%independent
random fields $D_i\in[-W,W]$,
%parameters $B=4\Jmax$, $\alpha=1.13$,
and 
%units with 
$\hbar=1$.
Initializing the spins in the
N\'eel state 
$|\uparrow\downarrow\cdots\uparrow\downarrow\rangle$,
the system exhibits MBL for 
disorder strengths beyond 
about $W=\Jmax$ \cite{smi16}.
As was noted in Ref. \cite{wu16},
the experimentally measured
Hamming distance $\HD$ from 
Ref. \cite{smi16} can be recovered 
as the expectation value of the 
observable $A:=(1-M)/2$ with
$M:=N^{-1}\sum_i(-1)^i\sigma_i^z$
(staggered magnetization).
%(``spin imbalance between even 
%and odd sites''). 

In Fig. 1, the experimental results 
are compared with our theoretical approximation
(\ref{70}), (\ref{40}) by introducing
the numerically determined energies 
$E_n$ of the Hamiltonian (\ref{90})
into Eq. (\ref{50}).
Furthermore, as in the experiment, 
we averaged the so 
obtained results for $\tilde D(t)$
over 30 realizations of the 
disorder in Eq. (\ref{90}).
Since there are only $N=10$ spins,
$\max_n\rho_{nn}(0)$ 
is typically not yet very 
small and increases with $W$.
%a rather small number of levels
%are found to exhibit relatively 
%large populations $\rho_{nn}(0)$, 
%%and the number of those dominant 
%%$\rho_{nn}(0)$ further 
%%which further decreases 
%%with increasing 
%%disorder strength 
%$W$.
%%Since the theory requires that
%%$\max_n \rho_{nn}(0)$ is small,
We therefore focused on a moderate 
disorder of $W=4\Jmax$, and we considered 
labels $n$ with $\rho_{nn}(0) < 0.01$
as negligible (see above),
resulting in typical values
$\max_n \rho_{nn}(0)\approx 0.1$
and $D\approx 20$.
The concomitant approximations for
$\langle A\rangle_{\!\rho(0)}$
turned out to exhibit particularly 
strong finite-$N$ effects; hence,
we used the {\em a priori}
known actual value 
$\langle A\rangle_{\!\rho(0)}=0$
in Eq. (\ref{70}).

Besides those disorder averaged 
results, individual 
realizations of Eq. (\ref{90}) 
would also seem interesting.
Since 
%such 
experimental data
are not available, we 
%instead
replicated the numerical solutions
of the Schr\"odinger equation
with Hamiltonian (\ref{90})
from Refs. \cite{smi16,wu16}.
The results for two realizations
are shown in the insets of Fig. 1.
The theoretical curves have been 
obtained as described above, employing 
the same realization of Eq. (\ref{90}) 
as in the numerics in each inset.
In view of those quite notable
finite size fluctuations,
the theory explains the 
%experimental and numerical
%observed 
``real'' temporal relaxation remarkably well.
%the theory agrees with the
%experiment and the numerics 
%as well as it possibly can
%(see below Eq. (\ref{10})).
%In particular, thermalization
%would imply that $\HD(t)$ equilibrates
%towards $1/2$.

\begin{figure}
\epsfxsize=1.0\columnwidth
\epsfbox{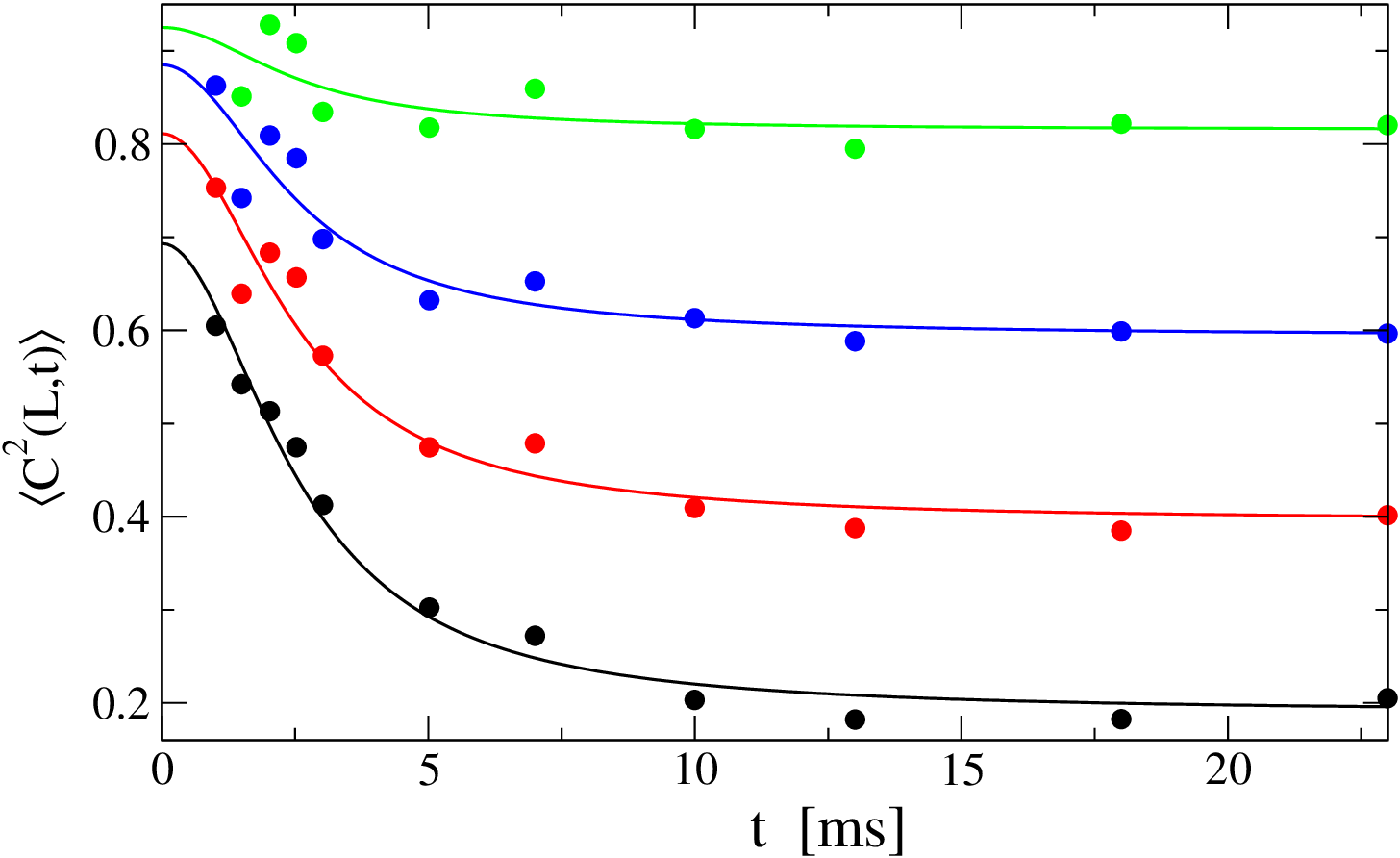}
\caption{\label{fig2}
Symbols: Experimental 
mean integrated squared contrast 
from Fig. 3 in Ref. \cite{kuh13}
for integration lengths 
$L=18,\,40,\,60,\,100\,\mu$m 
(from top to bottom) and vertically 
shifted by $0.3,\, 0.2,\, 0.1,\, 0.0$, 
respectively, for better visibility.
Lines:
Theoretical approximations (\ref{70}), (\ref{80}) with $T=3\,$nK.
%Theory from (\ref{70}) with
%(\ref{80}), $\tau:=\hbar/k_B T$, 
%and $T=3\,$nK.
%For further details see main text.
}
\end{figure}

Next, we consider the 
%temporal 
%relaxation 
equilibration
of a coherently split Bose gas, as
observed experimentally by Kuhnert et al.
in Ref. \cite{kuh13} via the mean integrated 
squared contrast $\langle C^2(L,t)\rangle$ 
of the matter-wave interference pattern 
for various integration lengths $L$.
This experiment (approximately) realizes 
an integrable system, 
%namelya Luttinger liquid, 
exhibiting prethermalization rather 
than thermalization \cite{gri12}.
The data from Ref. \cite{kuh13}
are compared in Fig. 2 with our
theory equation (\ref{70}).
Since modeling the quite intricate 
observable of the actual experiment
%$A$ corresponding to the actual 
%measurements $\langle C^2(L,t)\rangle$
%is clearly hopeless, 
goes beyond our present scope,
we treated 
%the expectation values
%for $t=0$ and $t\to\infty$, i.e.,
$\langle A\rangle_{\!\rho(0)}$
and $\langle A\rangle_{\!\rhobar}$
in Eq. (\ref{70})
%, 
as fit parameters for any given $L$.
Similarly, estimating 
the experimentally relevant 
``effective temperature'' $T$ in Eq. (\ref{80})
from first principles 
is beyond our present scope; hence,
%According to the discussion below
%(\ref{80}), we furthermore employed
%the approximation (\ref{80})
%We further assumed that if the system 
%{\em were} at thermal equilibrium 
%%{\em were} in a thermal equilibrium state
%%{\em would} thermalize 
%it obeyed standard ESM, hence we can 
%employ (\ref{80}) 
%with some suitable effective 
%temperature $T$.
%in $\tau:=\hbar/k_BT$.
%But since the system does not 
%thermalize for the initial 
%condition realized in the 
%experiment, it is again quite
%hopeless to theoretically 
%estimate this effective 
%temperature.
%We therefore used
%Since it is again beyond our present 
%scope to estimate its value from 
%first principles,
%we used $T$ as one more fit 
%$T$ in (\ref{80})
it
was treated as fit parameter 
(common to all $L$), yielding
$T=3\,$nK.
In fact, Fig. 5 in Ref. \cite{rau16} 
suggests
%implies 
that the experimental estimate 
$\Teff\approx 10\pm 3\,$nK from Fig. 2 
(at $t_e=0\,$ms) in Ref. \cite{gri12} 
%still applies to 
may also be a reasonable approximation 
in our case.
Here, $\Teff$ is yet another effective temperature,
which would agree with $T$ at thermal equilibrium, 
but may well be different from $T$ 
in our present case.
Furthermore, the experimental estimate 
%from Ref. \cite{gri12}
of $\Teff$ is based on a quite involved 
procedure \cite{gri12},
whose implicit premises may only be
approximately satisfied.
In conclusion, $T=3\,$nK seems still 
compatible with the experimental findings, 
and the resulting theoretical curves
%in Fig. 2 nicely agree with the data
%within the measurement uncertainties.
in Fig. 2 
%agree with the data within the measurement 
%uncertainties as well as they possibly can.
explain the main features of the data 
quite well.

\begin{figure}
\epsfxsize=1.0\columnwidth
\epsfbox{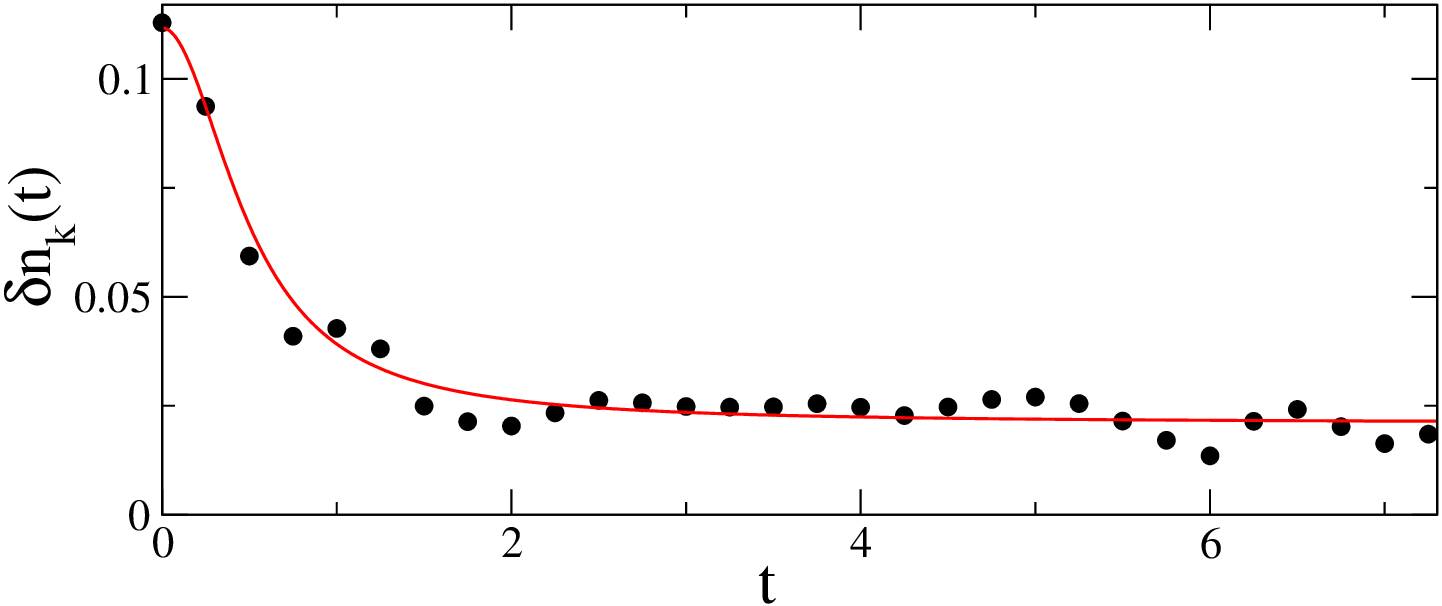}
\caption{\label{fig3}
Symbols: Numerical results from 
Fig 1(e) of Ref. \cite{rig09}.
Line: Theoretical approximation (\ref{70}), 
(\ref{80}).
% with $\tau=1/2$.
For further details, see the text.
}
\end{figure}

As a third example, we turn to the numerical 
results for an integrable model by Rigol \cite{rig09},
consisting of eight hard-core bosons on a 
periodic one-dimensional lattice with 24 sites,
and exhibiting 
%equilibration but not thermalization.
nonthermal longtime expectation values.
The detailed definition of the considered
observable $\delta n_k(t)$ from Ref. \cite{rig09}
is not repeated here since only the initial
and longtime values are actually
needed in Eq. (\ref{70}),
whose quantitative values cannot be
estimated theoretically anyway,
and hence are treated as fit parameters.
Furthermore, we adopted the approximation 
(\ref{80}) with the estimate 
%for the effective temperature 
$T=2$ from Ref. \cite{rig09} 
(in units with $\kB=\hbar=1$).
The resulting agreement with the numerical 
data in Fig. \ref{fig3} is remarkably good,
considering that the system consists of 
just eight bosons.

\begin{figure}
\epsfxsize=1.0\columnwidth
\epsfbox{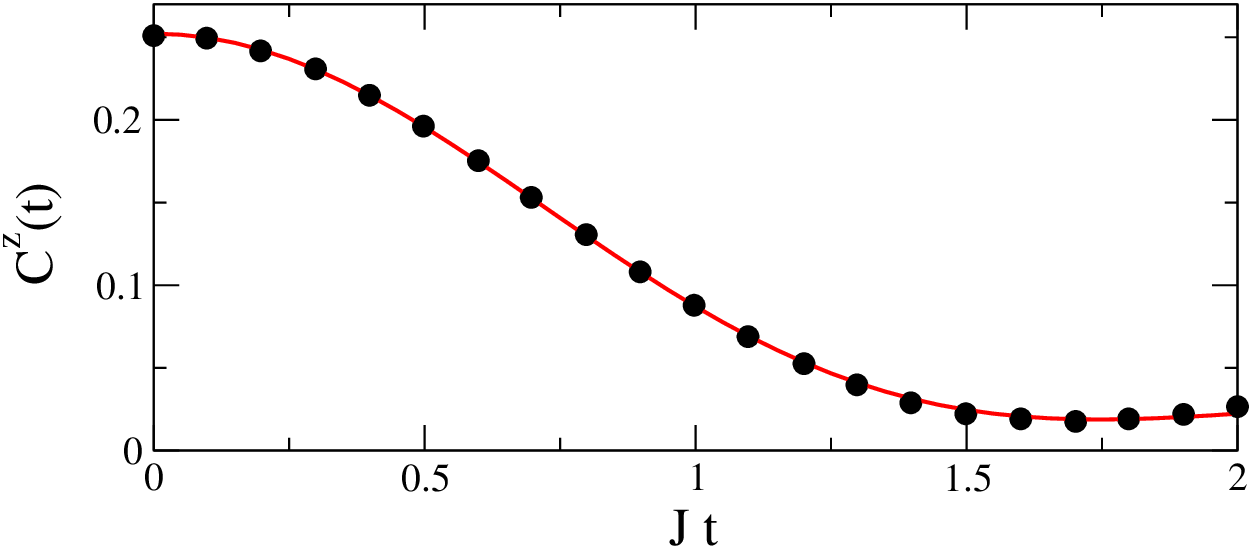}
\caption{\label{fig4}
Symbols: Numerical results 
for the spin-spin correlation 
%in the $z$ direction 
%$C_{8,9}^z(t)$ 
$C^z(t)$ 
from Fig. 8 of Ref. \cite{tor14}
for a spin-1/2 XXZ model with 
16 spins, coupling $J$,
anisotropy $\Delta=1/2$,
$\hbar=1$,
and a so-called 
pairs of parallel spins
initial condition.
Line: Theoretical approximation 
(\ref{70}) as specified in the 
text.
}
\end{figure}

Our last example is the integrable 
XXZ model of Torres-Herrera
et al. from Ref. \cite{tor14}.
Similarly as before, the 
initial value $\langle A\rangle_{\!\rho(0)}=0.25$
in Eq. (\ref{70}) is known {\em a priori} 
for the specific observable under 
consideration, while $\langle A\rangle_{\!\rhobar}$
is treated as a fit parameter.
On the other hand, $F(t)$ 
is now evaluated via Eq. (\ref{40})
by approximating the discrete
levels on the right-hand side 
of Eq. (\ref{50}) by a continuous 
level density \cite{rei16}.
In view of Table 1 and 
Fig 3(b) in Ref. \cite{tor14}, 
we roughly approximated this
density as constant
within the energy interval
$\IE=[-1.8,1.8]$ and as zero otherwise.
The resulting agreement with the 
numerics in Fig. \ref{fig4} speaks 
for itself.

In conclusion, we devise in this Letter a 
general analytical theory for the temporal 
relaxation behavior of isolated many-body 
systems which do not thermalize.
%The main prerequisites are that the initial state 
%exhibits no significant spatial inhomogeneities
%and notably populates many energy levels.
The main prerequisites are initial conditions 
which appreciably populate many energy 
levels and do not give rise to 
significant spatial inhomogeneities 
on macroscopic scales.
Specifically, the relaxation must 
not entail any significant transport 
currents, caused by some unbalanced local 
densities (of particles, energy, etc).
%Furthermore, the considered observable
%must not amount to a small perturbation
%of a conserved quantity.
On the other hand, the particular
reason for the absence of thermalization 
(MBL, integrability, broad energy 
distribution) seems largely irrelevant.
In fact, our theory also applies to 
systems which do thermalize, provided 
that the diagonal and microcanonical 
ensembles yield identical expectation 
values due to, e.g., the validity 
of the ETH.
Compared to previous related studies,
our main new concept consists in
admitting only permutations of basis vectors
in Eq. (\ref{20}), rather than arbitrary
(Haar distributed) basis transformations,
thus preserving all local constants 
of motion,
the diagonal ensemble which 
governs the longtime behavior,
and the violation (or not) 
of the ETH.
The adequate treatment of
inhomogeneous initial conditions 
%and of almost conserved observables
remains an important challenge for
future research.

\begin{acknowledgments}
We are indebted to Marcos Rigol for providing
the original data from Fig 1(e) of Ref. \cite{rig09}.
This work was supported by DFG Grants 
No. RE1344/7-1 and No. RE1344/10-1, and by the 
Studienstiftung des Deutschen Volkes.
\end{acknowledgments}

\newpage
\begin{center}
{\bf{\large{SUPPLEMENTAL MATERIAL}}}
\end{center}

\section{Averages over permutations}
As in the main text, we denote by $\Pi$ the set of
all permutations of $\{1,...,D\}$ for an arbitrary
but fixed $D\in\NN$. Hence, there are $D!$
elements $\pi\in\Pi$.

Next, we consider an arbitrary but fixed 
$n\in\{1,...,D\}$ and define
\begin{eqnarray}
\Pi_\nu := \{\pi\in\Pi\, | \, \pi(n)=\nu \}
\label{s1}
\end{eqnarray}
for any $\nu\in\{1,...,D\}$.
It follows that every $\pi\in\Pi$ belongs
to one and only one of those subsets 
$\Pi_\nu\subset\Pi$.
For symmetry reasons, each of the $D$ subsets 
$\Pi_\nu$ contains the same number of elements, 
and hence this number must be $(D-1)!$.

Denoting, as in the main text, by $[...]_\Pi$
the average over all $\pi\in\Pi$, we can conclude 
for an arbitrary but fixed 
$n\in\{1,...,D\}$ and any function 
$f:\{1,...,D\}\to\CC$ that
\begin{eqnarray}
\!\!\! [f(\pi(n))]_\Pi := \frac{1}{D!}\sum_{\pi\in\Pi}f(\pi(n))
= \frac{1}{D!}\sum_{\nu=1}^D \sum_{\pi\in\Pi_\nu}
\!f(\pi(n))
\label{s2}
\end{eqnarray}
where the last identity is due to the above observation 
that $\Pi$ is the disjoint union of all the $\Pi_\nu$'s.
Considering that $\pi(n)=\nu$ for all $\pi\in\Pi_\nu$
according to (\ref{s1}), we can conclude that
\begin{eqnarray}
[f(\pi(n))]_\Pi 
= 
\frac{1}{D!}\sum_{\nu=1}^D (D-1)! f(\nu)
=
\frac{1}{D}\sum_{\nu=1}^D f(\nu)
\label{s3}
\end{eqnarray}
independent of $n$.

Analogously, we consider an arbitrary but fixed 
pair of integers $m,n\in\{1,...,D\}$ 
with $m\not=n$
and define
\begin{eqnarray}
\Pi_{\mu\nu} := \{\pi\in\Pi\, | \, \pi(m)=\mu,\ \pi(n)=\nu \}
\label{s4}
\end{eqnarray}
for any $\mu,\, \nu\in\{1,...,D\}$ with $\mu\not=\nu$.
As before, it follows that every $\pi\in\Pi$ is contained
to one and only one 
%of those subsets 
$\Pi_{\mu\nu}$
%\subset\Pi$ 
and that every
$\Pi_{\mu\nu}$ consists of $(D-2)!$ elements.
For an arbitrary complex values function
$f$ of two arguments $m,n\in \{1,...,D\}$
%$f:\{1,...,D\}\times\{1,...,D\}\to\CC$ 
we thus obtain
\begin{eqnarray}
& &
[f(\pi(m),\pi(n))]_\Pi :=  \frac{1}{D!}\sum_{\pi\in\Pi}f(\pi(m),\pi(n))
\nonumber
\\
& & 
=  \frac{1}{D!}\, {\sum_{\mu\nu}}'\!
\sum_{\pi\in\Pi_{\mu\nu}}
f(\pi(m),\pi(n))
\ .
\label{s5}
\end{eqnarray}
Here and in the following, the prime symbol in 
$\sum'$ means that all summation indices 
run from $1$ to $D$ and must be pairwise distinct.
As before, one now can infer that
\begin{eqnarray}
[f(\pi(m),\pi(n))]_\Pi
& = &
\frac{1}{D!}{\sum_{\mu\nu}}' (D-2)! f(\mu,\nu)
\nonumber
\\
& = &
\frac{1}{D(D-1)}{\sum_{\mu\nu}}' f(\mu,\nu)
\label{s6}
\end{eqnarray}
for any $m\not=n$.

A straightforward generalization of the above line
of reasoning yields
\begin{eqnarray}
\![f(\pi(n_1),...,\pi(n_K))]_\Pi 
= \frac{(D-K)!}{D!}\!{\sum_{\nu_1...\nu_K\!\!\!\!\!\!\!}}'
f(\nu_1,...,\nu_K) 
\label{s7}
\end{eqnarray}
for any $K$-tuple $(n_1,...,n_K)$ of pairwise distinct 
integers $n_k\in\{1,...,D\}$ and any complex valued 
function $f(n_1,...,n_K)$.

%%%%%%%%%%%%%%%%%%%%%%%%%%%%%%%%%%%%%%%%%%%
\section{Subsets of quadruples}
For any given $D\in\NN$ we define the set
of quadruples 
\begin{eqnarray}
I:=\{\, (k,l,m,n)\ |\ k,l,m,n\in\{1,...,D\}\, \}\ .
\label{t1}
\end{eqnarray}
Next, we introduce 15 subsets
$I_a$ of $I$, $a=1,...,15$,
defined via the following
properties of its elements $(k,l,m,n)$:
\begin{eqnarray}
I_1 & : & k=l=m=n
\nonumber
\\
I_2 & : & l=m=n,\ k\not=l
\nonumber
\\
I_3 & : & k=m=n,\ l\not=k
\nonumber
\\
I_4 & : & k=l=n,\ m\not=n
\nonumber
\\
I_5 & : & k=l=m,\ n\not=m
\nonumber
\\
I_8 & : & k=l,\ m=n,\ k\not=m
\nonumber
\\
I_9 & : & k=m,\ l=n,\ k\not=l
\nonumber
\\
I_{10} & : & k=n,\ l=m,\ k\not=l
\nonumber
\\
I_6 & : & m=n\ \mbox{and $k,l,m$ pairwise distinct}
\nonumber
\\
I_7 & : & k=l\ \mbox{and $k,m,n$ pairwise distinct}
\nonumber
\\
I_{11} & : & k=m\ \mbox{and $k,l,n$ pairwise distinct}
\nonumber
\\
I_{12} & : & l=n\ \mbox{and $k,l,m$ pairwise distinct}
\nonumber
\\
I_{13} & : & l=m\ \mbox{and $k,l,n$ pairwise distinct}
\nonumber
\\
I_{14} & : & k=n\ \mbox{and $k,l,m$ pairwise distinct}
\nonumber
\\
I_{15} & : & \mbox{$k,l,m,n$ pairwise distinct}
\label{t2}
\end{eqnarray}
The sequence of labels of $I_6,...,I_{10}$ may 
appear peculiar but will turn out be convenient 
later on. It is quite obvious that every given
quadruple $(k,l,m,n)\in I$ is contained in
one and only one of the 15 subsets $I_a$:
Either all indices $k,l,m,n$ are equal ($I_1$),
or three of them are equal and the 
fourth distinct ($I_2,...,I_5$),
or they can be grouped into two distinct pairs
($I_8,I_9,I_{10})$,
or three are pairwise distinct and 
one among them is equal to the fourth
($I_6,I_7,I_{11},...,I_{14})$,
or all of them are pairwise distinct ($I_{15}$).
%One readily can convince oneself that every
%quadruple $(k,l,m,n)\in I$ is contained in
%one and only one of the 15 subsets $I_a$.
In other words, $I$ is the disjoint union
of the 15 subset $I_a$.

Finally, we consider the set of quadruples
\begin{eqnarray}
J:=\{\, (k,l,m,n)\in I\ | \ k\not=l,\ m\not=n\, \}
\label{t3}
\end{eqnarray}
Since $J\subset I$ and since $I$ is the disjoint 
union of the 15 subsets $I_a$, it follows that
J is the disjoint union of the 15 subsets
$J_a:=J\cap I_a$.
The definitions (\ref{t2}) and (\ref{t3})
readily imply that the subsets $J_1,...,J_8$
are empty
%=\{\}$ for $a=1,...,8$
and that $J_a=I_a$ for $a=9,...,15$.
In conclusion $J$ is the disjoint union
of the subsets $I_a$ with $a=9,...,15$.

%%%%%%%%%%%%%%%%%%%%%%%%%%%%%%%%%%%%%%%%%%%
\section{Derivation of the main result}
\subsection{Preliminaries}
As in the main text, we consider an
arbitrary but fixed Hamiltonian
$H_1=\sum_{n=1}^D E_n |n\rangle\langle n|$.
Furthermore, we recall the definition 
from Eq. (2) in the main text,
namely
\begin{eqnarray}
H_\pi := \sum_{n=1}^D E_n |\pi(n)\rangle\langle \pi(n)|
\ .
\label{u1}
\end{eqnarray}
%with $n_\pi:=\pi(n)$.
%and where $\pi^{-1}$ denotes the inverse of 
%any given permutation $\pi\in\Pi$.
Similarly as around Eq. (1) of the main text,
the time evolution induced by $H_\pi$
is $\rho(t)=\propa\rho(0)\propa^\dagger$
with $\propa:=e^{-\i H_\pi t/\hh}$
and the expectation value
$\langle A\rangle_{\!\rho(t)}:= \tr\{\rho(t) A\}$
takes the form
\begin{eqnarray}
\langle A\rangle_{\!\rho(t)}
& = &  \sum_{m,n=1}^D
g(\pi(m),\pi(n))\ e^n_m
\label{u2}
\\
g(m,n) & := & \rho_{mn}(0)\, A_{nm}
\label{u3}
\\
e^n_m & := & e^{\i (E_n-E_m)t/\hh} 
\ ,
\label{u4}
\end{eqnarray}
where, for notational convenience, the $t$ 
dependence of $e_m^m$ is omitted, and 
where, as in the main text,
$A_{mn}:=\langle m |A| n \rangle$,
$\rho_{mn}(t):=\langle m|\rho(t)|n\rangle$.

Denoting, as usual, by $[...]_\Pi$ the average over
all $\pi\in\Pi$, the first objective of the following
calculations is to explicitly evaluate the average
$[\langle A\rangle_{\!\rho(t)}]_\Pi$
of the expectation value from (\ref{u2}).
Denoting the deviation from the average by
\begin{eqnarray}
\xi_\pi(t) & := & \langle A\rangle_{\!\rho(t)}-
[\langle A\rangle_{\!\rho(t)}]_\Pi \ ,
\label{u4a}
\end{eqnarray}
our second objective is to show that this 
definition implies (3) of the main
text and that the variance of
$\xi_\pi(t)$ satisfies the second 
relation in (6) from the main text.
Note that the first relation in (6)
follows immediately from the definition
(\ref{u4a}).

Before starting with those calculations,
we introduce some useful definitions and 
relations.
As in the main text, the diagonal ensemble 
$\rhobar$ is defined via 
its matrix elements as
\begin{eqnarray}
\rhobar_{mn} := \delta_{mn}\, \rho_{nn}(0) \ ,
\label{u4b}
\end{eqnarray}
where $\delta_{mn}$ is the Kronecker delta.
One readily sees that $\rhobar$ is hermitian,
non-negative and of unit trace, i.e.
a well-defied density operator.
Likewise, we introduce the auxiliary
hermitian operators $B$ and $C$ via
\begin{eqnarray}
B_{mn} :=  \delta_{mn}\, A_{nn} \ ,
\ \ 
C_{mn} := A_{mn}-B_{mn}
\ .
\label{u4d}
\end{eqnarray}
Denoting by $\norm{A}$ the operator norm of
$A$, i.e. the largest eigenvalue in modulus,
one readily concludes that
\begin{eqnarray}
\norm{B} \leq \norm{A} \ , \ \ 
\norm{C}  \leq &  2 \norm{A} \ .
\label{u4f}
\end{eqnarray}

As in the main text, we denote by $\da$ 
the difference between the largest and smallest 
eigenvalues of $A$.
It is intuitively obvious, and can also be
readily verified rigorously, that
$\xi_\pi(t)$ from (\ref{u4a}) remains unchanged 
upon adding an arbitrary constant to $A$.
Without loss of generality, we thus can
assume that the largest and smallest eigenvalues
of $A$ are of equal modulus and opposite sign.
Hence, we can take for granted that
\begin{eqnarray}
\norm{A} = \da/2
\label{u4g}
\end{eqnarray}
as far as the properties of $\xi_\pi(t)$ 
are concerned.

Denoting the eigenvalues and eigenvectors
of $\rho$ as $p_n$ and $|\chi_n\rangle$,
respectively, and observing that $p_n\geq 0$
for all $n=1,...,D$, it follows that
$\sqrt{\rho}:=\sum_{n=1}^D \sqrt{p_n}\,
|\chi_n\rangle\langle\chi_n|$ is a well-defined Hermitian operator with $(\sqrt{\rho})^2=\rho$.
Given an arbitrary but fixed pair of vectors
$|\psi\rangle$ and $|\phi\rangle$ we define
$|\psi'\rangle:=\sqrt{\rho}\,|\psi\rangle$ and 
$|\phi'\rangle:=\sqrt{\rho}\,|\phi\rangle$.
Rewriting $\langle\psi|\rho|\phi\rangle$
as $\langle\psi'|\phi'\rangle$ and
invoking the Cauchy-Schwarz inequality,
we can conclude that
\begin{eqnarray}
|\langle\psi|\rho|\phi\rangle|^2\leq
\langle\psi|\rho|\psi\rangle\langle\phi|\rho|
\phi\rangle \ .
\label{u4m}
\end{eqnarray}

Finally, we recall the definitions (4) and 
(5) from the main text, reading
\begin{eqnarray}
F(t) & := & \left(\DD \,|\phi(t)|^2-1\right)/(D-1) \ ,
\label{u4h}
\\
\phi(t) & := & 
\DD^{-1}\mbox{$\sum_{n=1}^{\DD}$}e^{\i E_n t/\hh} 
\ .
\label{u4i}
\end{eqnarray}
By exploiting (\ref{u4}) one readily concludes that
\begin{eqnarray}
F(t) 
 & = & \frac{1}{D(D-1)}
\left[ \sum_{m,n=1}^D e^{\i (E_n-E_m) t/\hh} - D \right]
\nonumber
\\
& = &  
\frac{1}{D(D-1)} {\sum_{mn}}' e^n_m
\ .
\label{u4k}
\end{eqnarray}
Here and in the following, and in accordance
with the notation in (\ref{s5})-(\ref{s7}),  
the prime symbol in $\sum'$ indicates 
that all summation indices run from 
$1$ to $D$ and must be pairwise distinct.

%%%%%%%%%%%%%%%%%%%%%%%%%%%%%%%%%%%%
\subsection{Evaluation of the average}
In view of (\ref{u2}), we can infer that
\begin{eqnarray}
[\langle A\rangle_{\!\rho(t)}]_\Pi
\!\! & = &  \!\! \sum_{m,n=1}^D
[g(\pi(m),\pi(n))]_\Pi\ e^n_m = Q+R
\label{u5}
\\
Q & := & \sum_{n=1}^D\ [g(\pi(n),\pi(n))]_\Pi
\label{u6}
\\
R & := & {\sum_{mn}}' \, [g(\pi(m),\pi(n))]_\Pi\ e^n_m
\ .
\label{u7}
\end{eqnarray}
In (\ref{u6}) we exploited that $e^n_n=1$ 
according to (\ref{u4}), and the
primed sum in (\ref{u7}) is defined 
below (\ref{u4k}).
Choosing $f(n):=g(n,n)$ in (\ref{s3}), 
we can conclude that
\begin{eqnarray}
Q = \sum_{\nu=1}^D g(\nu,\nu) 
\label{u8}
\end{eqnarray}
and choosing $f(m,n):=g(m,n)$ in (\ref{s6}) that
\begin{eqnarray}
R = {\sum_{mn}}' e^n_m \frac{1}{D(D-1)}
{\sum_{\mu\nu}}' g(\mu,\nu)
\ .
\label{u9}
\end{eqnarray}
Taking into account (\ref{u4k}) 
and (\ref{u8}), the last relation can
be rewritten as
\begin{eqnarray}
R = F(t)
\left[\sum_{\mu,\nu=1}^D g(\mu,\nu) -Q\right] 
\ .
\label{u9a}
\end{eqnarray}
The definition (\ref{u3}) implies that
\begin{eqnarray}
\sum_{\mu,\nu=1}^D g(\mu,\nu)
& = & \sum_{\mu,\nu=1}^D 
\langle \mu | \rho(0) |\nu \rangle
\langle \nu | A |\mu \rangle
\nonumber
\\
& = & 
\sum_{\mu=1}^D 
\langle \mu | \rho(0) A |\mu \rangle
\nonumber
\\
& = & 
%\tr\{\rho(0)A\}=\langle A\rangle_{\!\rho(0)}
\tr\{\rho(0)A\}
\ .
\label{u9b}
\end{eqnarray}
Likewise, one readily sees that
\begin{eqnarray}
\sum_{\nu=1}^D g(\nu,\nu)
& = & 
%\langle A\rangle_{\!\rhobar}
\tr\{\rhobar A\}
\ .
\label{u9c}
\end{eqnarray}
By introducing (\ref{u9b}) and (\ref{u9c}) 
into (\ref{u8}) and (\ref{u9a}),
one finds that
\begin{eqnarray}
Q & = & \tr\{\rhobar A\}
\label{u9e}
\\
R & = & F(t)[\tr\{\rho(0)A\}-\tr\{\rhobar A\}]
\ .
\label{u9f}
\end{eqnarray}
Adopting the definition
$\langle A\rangle_{\!\rho}:= \tr\{\rho A\}$
from the main text, we finally 
can rewrite (\ref{u5}) as
\begin{eqnarray}
[\langle A\rangle_{\!\rho(t)}]_\Pi
& = & 
\langle A\rangle_{\!\rhobar} 
+
F(t)\, \left\{ \langle A\rangle_{\!\rho(0)} 
- \langle A\rangle_{\!\rhobar}\right\}
\label{u9d}
\end{eqnarray}
and with (\ref{u4a}) we recover
Eq. (3) from the main text.
%as well as the first relation 
%in (6).

For later use, we employ (\ref{u4b}) and
(\ref{u4d}) to conclude
\begin{eqnarray}
\tr\{\rhobar A\} 
=
\sum_{n=1}^D \rho_{nn}(0) A_{nn}
=\tr\{\rho(0) B\}
\label{u9g}
\end{eqnarray}
and hence
\begin{eqnarray}
\tr\{\rho(0)A\}-\tr\{\rhobar A\}
 & = &
\tr\{\rho(0) (A-B)\}
\nonumber
\\
& = &
\tr\{\rho(0) C\}
\ .
\label{u9h}
\end{eqnarray}
With (\ref{u4k}) we thus can rewrite (\ref{u9f}) as
\begin{eqnarray}
R = \tr\{\rho(0) C\} \frac{1}{D(D-1)} {\sum_{mn}}' e^n_m
\ .
\label{u9i}
\end{eqnarray}

%%%%%%%%%%%%%%%%%%%%%%%%%%%%%%%%%%%%
\subsection{Evaluation of the variance}
From (\ref{u2})-(\ref{u4}) we can infer that
\begin{eqnarray}
\!\!\!\! & & \!\!\!\! (\langle A\rangle_{\!\rho(t)})^2
=\!\! \sum_{klmn\in I} 
\!\!
h(\pi(k),\pi(l),\pi(m),\pi(n))
\ e^{ln}_{km}
\ ,
\label{u10}
\end{eqnarray}
where
\begin{eqnarray}
h(k,l,m,n) & := & g(k,l)g(m,n)
\label{u11}
\\
%\!\!\!\! & & \!\!\!\! 
%\rho_{mn} := \rho_{mn}(0)
%\label{u12}
%\\
e^{ln}_{km} & := & 
e^l_k\, e^n_m
%e^{\i(E_l-E_k+E_n-E_m)t/\hh}
\label{u13} \ .
\end{eqnarray}
In (\ref{u10}), the symbol $\sum_{klmn\in I}$ indicates
a summation over all quadruples of indices 
contained in $I$ from (\ref{t1}).
As pointed out below (\ref{t2}), this set $I$ is
the disjoint union of the 15 subsets $I_a$.
We thus can conclude from (\ref{u10}) that
\begin{eqnarray}
& & 
[(\langle A\rangle_{\!\rho(t)})^2]_\Pi
= \sum_{a=1}^{15} S_a
\label{u14}
\\
& & 
S_a := \!\! \sum_{klmn\in I_a} 
\!\!
[h(\pi(k),\pi(l),\pi(m),\pi(n))]_\Pi \ e^{ln}_{km} \ .
\label{u15}
\end{eqnarray}

Next, we turn to the evaluation 
of the 15 terms $S_a$ from (\ref{u15}).
To begin with, one readily sees 
that $I_2$ goes over into $I_4$ 
in (\ref{t2}) upon interchanging
the labels $k$ and $m$ and simultaneously 
interchanging the labels $l$ and $n$.
Similarly, $I_3$ goes over into
$I_5$ and $I_6$ into $I_7$.
On the other hand, the terms
(\ref{u11}) and (\ref{u13}) and hence 
the summands in (\ref{u15}) are invariant 
under such an interchange of labels. 
It follows that
\begin{eqnarray}
S_4 & = & S_2
\label{u16}
\\
S_5 & = & S_3
\label{u17}
\\
S_7 & = & S_6 \ .
\label{u18}
\end{eqnarray}
Likewise, upon interchanging
the labels $k$ and $l$ and simultaneously 
the labels $m$ and $n$, the subset
$I_{11}$ in (\ref{t2}) goes over 
into $I_{12}$ and the subset 
$I_{13}$ into $I_{14}$.
Furthermore, in view of (\ref{u3}) 
and (\ref{u4}) one can conclude that the terms
(\ref{u11}) and (\ref{u13}) and hence 
the summands in (\ref{u15}) go over into
their complex conjugate under such an 
interchange of labels.
It follows that
\begin{eqnarray}
S_{12} & = & S_{11}^\ast
\label{u19}
\\
S_{14} & =
 & S_{13}^\ast \ .
\label{u20}
\end{eqnarray}

%%%%%%%%%%%%%%%%%%%%%%%%%%%%%%%%%%%%
\subsection{Evaluation of $S_1,...,S_8$}
For $a=1$, all four indices in (\ref{t2}) 
are equal.
With (\ref{u4}) it follows that
the term (\ref{u13}) is unity 
and (\ref{u15}) takes the form
\begin{eqnarray}
S_1 & = & \sum_{n=1}^D [h(\pi(n),\pi(n),\pi(n),\pi(n))]_\Pi
\ .
\label{u21}
\end{eqnarray}
Choosing $f(n):=h(n,n,n,n)$ in (\ref{s3}), 
we thus obtain
\begin{eqnarray}
S_1 & = & \sum_{\nu=1}^D h(\nu,\nu,\nu,\nu)
\ .
\label{u22}
\end{eqnarray}

Along the same line of reasoning one finds for 
$a=8$ by choosing $f(m,n):=h(m,m,n,n)$ 
that
in (\ref{s6}) that
\begin{eqnarray}
S_8 & = & {\sum_{ln}}' [h(\pi(l),\pi(l),\pi(n),\pi(n))]_\Pi
\nonumber
\\
& = & 
{\sum_{\lambda\nu}}' 
h(\lambda,\lambda,\nu,\nu)
\ .
\label{u23}
\end{eqnarray}
Combining (\ref{u22}) and (\ref{u23})
yields
\begin{eqnarray}
S_1+ S_8 & = & \sum_{\lambda\nu=1}^D 
h(\lambda,\lambda,\nu,\nu) 
\ .
\label{u24}
\end{eqnarray}
%Comparing (\ref{u3}) with (\ref{u11}) implies that
%$g(\lambda,\lambda,\nu,\nu)=f(\lambda,\lambda)f(\nu,\nu)$.
Observing (\ref{u11}) and (\ref{u8}) 
we thus can rewrite (\ref{u24}) as
\begin{eqnarray}
S_1+S_8 = Q^2 \ .
\label{u25}
\end{eqnarray}

Along the same line of reasoning one finds for $a=6$ that
\begin{eqnarray}
S_6 & = & {\sum_{kln}}' 
[h(\pi(k),\pi(l),\pi(n),\pi(n))]_\Pi \ e^{ln}_{kn}
\nonumber
\\
& = & 
{\sum_{kln}}'e^{ln}_{kn}\frac{1}{D(D-1)(D-2)}
{\sum_{\kappa\lambda\nu}}' 
h(\kappa,\lambda,\nu,\nu)
\ .
\label{u26}
\end{eqnarray}
In the last step, we exploited (\ref{s7}) 
with $K=3$ and $f(k,l,n):=g(k,l,n,n)$.
Due to (\ref{u13}) and (\ref{u4}) it follows that
$e^{ln}_{kn}=e^l_k$, hence the summands in
(\ref{u26}) are independent of $n$.
Since the prime indicates that all the
summands $k,l,n$ must be pairwise distinct,
it follows that for every given pair $(k,l)$,
the index $n$ can take on $D-2$ different
values. Performing the summation over $n$
thus yields a factor $D-2$ and we obtain
\begin{eqnarray}
S_6 & = & 
{\sum_{kl}}'e^{l}_{k}\frac{1}{D(D-1)}
{\sum_{\kappa\lambda\nu}}' 
h(\kappa,\lambda,\nu,\nu)
\ .
\label{u27}
\end{eqnarray}

The evaluation of $S_2$ and $S_3$ is similar to
(but simpler than) that of $S_6$, yielding
\begin{eqnarray}
S_2 & = & 
{\sum_{kl}}'e^{l}_{k}\frac{1}{D(D-1)}
{\sum_{\kappa\lambda}}' 
h(\kappa,\lambda,\lambda,\lambda)
\label{u28}
\\
S_3 & = & 
{\sum_{kl}}'e^{l}_{k}\frac{1}{D(D-1)}
{\sum_{\kappa\lambda}}' 
h(\kappa,\lambda,\kappa,\kappa)
\label{u29}
\end{eqnarray}

Altogether (\ref{u27})-(\ref{u29}) sum up to
\begin{eqnarray}
\!\!\! 
S_2+S_3+S_6 \! = \!
{\sum_{kl}}'e^{l}_{k}\frac{1}{D(D-1)}
{\sum_{\kappa\lambda}}'\sum_{\nu=1}^D 
h(\kappa,\lambda,\nu,\nu)
\ .
\label{u30}
\end{eqnarray}
Considering (\ref{u11}) and exploiting 
(\ref{u8}), (\ref{u9}) we finally can conclude that
\begin{eqnarray}
S_2+S_3+S_6 & = & Q\, R
\label{u31}
\end{eqnarray}
and with (\ref{u16})-(\ref{u18}), (\ref{u25}) that
\begin{eqnarray}
\sum_{a=1}^8 S_a& = & Q^2 + 2\, Q\, R
\ .
\label{u32}
\end{eqnarray}

%%%%%%%%%%%%%%%%%%%%%%%%%%%%%%%%%%%%
\subsection{Upper bounds for $S_9,...,S_{14}$}
In view of $I_9$ in (\ref{t2}), the
sum $S_9$ from (\ref{u15}) takes the form
\begin{eqnarray}
S_9 & = & {\sum_{mn}}' 
[h(\pi(m),\pi(n),\pi(m),\pi(n))]_\Pi \ e^{nn}_{mm}
\nonumber
\\
& = & 
{\sum_{mn}}'e^{nn}_{mm}\frac{1}{D(D-1)}
{\sum_{\mu\nu}}' 
h(\mu,\nu,\mu,\nu)
\ ,
\label{u33}
\end{eqnarray}
where we exploited (\ref{s6}) 
with $f(m,n):=g(m,n,m,n)$
in the last step.
With (\ref{u13}) and (\ref{u4}) 
it follows that $|e^{nn}_{mm}|=1$.
Since the primed sum over $m$ and $n$
consists of $D(D-1)$ summands, we can
conclude that
\begin{eqnarray}
|S_9| & \leq & 
{\sum_{mn}}'|e^{nn}_{mm}|\frac{1}{D(D-1)}
{\sum_{\mu\nu}}' 
|h(\mu,\nu,\mu,\nu)|
\nonumber
\\
& = &
{\sum_{\mu\nu}}' 
|\rho_{\mu\nu}|^2|A_{\mu\nu}|^2
\ .
\label{u34}
\end{eqnarray}
In the last relation, we utilized (\ref{u3}) and (\ref{u11}),
and we adopted the abbreviation 
\begin{eqnarray}
\rho_{mn}:=\rho_{mn}(0) \ .
\label{u34a}
\end{eqnarray}
The Cauchy-Schwarz inequality 
(\ref{u4m}) implies
$|\rho_{mn}|^2\leq \rho_{mm}\rho_{nn}$.
Upon extending the sum in (\ref{u34})
over all $\mu$ and $\nu$, we thus obtain
\begin{eqnarray}
|S_9| & \leq & 
\sum_{\mu,\nu=1}^D 
\rho_{\mu\mu}\rho_{\nu\nu}|A_{\mu\nu}|^2
\ .
\label{u35}
\end{eqnarray}

Denoting by $a$ and $b$ two arbitrary $D\times D$
matrices with matrix elements $a_{\mu\nu}$ and 
$b_{\mu\nu}$, the definition
$(a,b):=\sum_{\mu,\nu=1}^D 
a^\ast_{\mu\nu}b_{\mu\nu}\, 
(|A_{\mu\nu}|^2+\epsilon)$ amounts to
a well-defined scalar product for any 
$\epsilon>0$.
Invoking the Cauchy-Schwarz inequality
and then letting $\epsilon\to 0$ we 
can conclude that
\begin{eqnarray}
\left|\sum_{\mu,\nu=1}^D \!
a^\ast_{\mu\nu}b_{\mu\nu}\, 
|A_{\mu\nu}|^2\right|^2
\! \leq \!
\sum_{\mu,\nu=1}^D 
\!
|a^\ast_{\mu\nu}|^2
|A_{\mu\nu}|^2
\sum_{\mu,\nu=1}^D 
\!
|b^\ast_{\mu\nu}|^2
|A_{\mu\nu}|^2
\nonumber
\end{eqnarray}
For the particular choice 
$a_{\mu\nu}:=\rho_{\mu\mu}$ (independent of $\nu$) and
$b_{\mu\nu}:=\rho_{\nu\nu}$ (independent of $\mu$),
the right-hand side of (\ref{u35}) can 
thus be further bounded to yield
\begin{eqnarray}
|S_9| & \leq & \sqrt{T\, U}
\label{u36}
\\
T & := & \sum_{\mu,\nu=1}^D \rho^2_{\mu\mu} |A_{\mu\nu}|^2
\label{u37}
\\
U & := & \sum_{\mu,\nu=1}^D \rho^2_{\nu\nu} |A_{\mu\nu}|^2
\label{u38}
\end{eqnarray}
Since $|A_{\mu\nu}|^2=
\langle\mu|A|\nu\rangle\langle\nu|A|\mu\rangle$,
the sum over $\nu$ in (\ref{u37}) can be readily 
performed,
yielding
\begin{eqnarray}
T & = & \sum_{\mu=1}^D \rho^2_{\mu\mu} \langle\mu|A^2|\mu\rangle
\label{u39}
\end{eqnarray}
The last factor can be estimated from above by 
$\norm{A^2}$, which in turn is equal to $\norm{A}^2$.
The remaining sum over $\mu$ can be identified with
$\tr\rhobar^2$ by exploiting (\ref{u34a}) and (\ref{u4b}).
Altogether, we thus obtain
\begin{eqnarray}
T & \leq & \norm{A}^2\tr\rhobar^2 \ .
\label{u40}
\end{eqnarray}
The same estimate readily carries over to $U$ from
(\ref{u38}), implying for (\ref{u36}) that
\begin{eqnarray}
|S_9| & \leq & \norm{A}^2\tr\rhobar^2 \ .
\label{u41}
\end{eqnarray}

Along the same line of reasoning,
one finds for $a=10$ that
\begin{eqnarray}
S_{10} & = & {\sum_{mn}}' 
[h(\pi(n),\pi(m),\pi(m),\pi(n))]_\Pi \ e^{mn}_{nm}
\nonumber
\\
& = & 
{\sum_{mn}}'e^{mn}_{nm}\frac{1}{D(D-1)}
{\sum_{\mu\nu}}' 
h(\mu,\nu,\nu,\mu)
\ .
\label{u42}
\end{eqnarray}
With (\ref{u13}) and (\ref{u4}) it follows that
$e^{mn}_{nm}=1$.
Since the primed sum over $m$ and $n$
consists of $D(D-1)$ summands, we can
conclude with
(\ref{u3}) and (\ref{u11}) that
\begin{eqnarray}
S_{10} & = & 
{\sum_{\mu\nu}}' 
|\rho_{\mu\nu}|^2|A_{\mu\nu}|^2
\ .
\label{u43}
\end{eqnarray}
The last sum is identical to that in (\ref{u34}).
Hence, one finds exactly as in 
(\ref{u34a})-(\ref{u41}) the estimate
\begin{eqnarray}
|S_{10}| & \leq & \norm{A}^2\tr\rhobar^2 \ .
\label{u44}
\end{eqnarray}

Likewise, one finds for $a=11$ that
\begin{eqnarray}
S_{11} & = & {\sum_{kln}}' 
[h(\pi(k),\pi(l),\pi(k),\pi(n))]_\Pi \ e^{ln}_{kk}
\nonumber
\\
& = & 
{\sum_{kln}}'e^{ln}_{kk}\frac{1}{D(D-1)(D-2)}
{\sum_{\kappa\lambda\nu}}' 
h(\kappa,\lambda,\kappa,\nu)
\ .
\label{u45}
\end{eqnarray}
Since $|e^{ln}_{kk}|=1$
and the primed sum over $k,l,n$
consists of $D(D-1)(D-2)$ summands,
we can conclude that
\begin{eqnarray}
|S_{11}| & \leq & 
|{\sum_{\kappa\lambda\nu}}' 
h(\kappa,\lambda,\kappa,\nu)|
\nonumber
\\
& = & |{\sum_{\kappa\lambda\nu}}'
\rho_{\kappa\lambda} A_{\lambda\kappa}
\rho_{\kappa\nu} A_{\nu\kappa}|
\ .
\label{u46}
\end{eqnarray}
In the last step, we utilized (\ref{u3}), (\ref{u11}),
and (\ref{u34a}).
The primed sum extends over all 
$\kappa,\lambda,\nu$
under the constraint that they 
must be pairwise distinct.
It can be rewritten as a sum over
all $\kappa,\lambda,\nu$ if we multiply 
each summand by an extra factor
$(1-\delta_{\lambda\kappa})
(1-\delta_{\nu\kappa})(1-\delta_{\nu\lambda})$.
Indeed, the latter factor is unity if 
$\kappa,\lambda,\nu$ are pairwise distinct, 
and zero otherwise.
Observing that 
$(1-\delta_{\lambda\kappa})
 A_{\lambda\kappa}
=C_{\lambda\kappa}$
and
$(1-\delta_{\nu\kappa})
 A_{\nu\kappa}
=C_{\nu\kappa}$
according to (\ref{u4d}),
we can conclude that
\begin{eqnarray}
|S_{11}| & \leq & 
|\sum_{\kappa\lambda\nu=1}^D 
(1-\delta_{\nu\lambda})
\rho_{\kappa\lambda} C_{\lambda\kappa}
\rho_{\kappa\nu} C_{\nu\kappa}|
\nonumber
\\
& \leq & V+W
\label{u47}
\\
V & := & |\sum_{\kappa\lambda\nu=1}^D 
\rho_{\kappa\lambda} C_{\lambda\kappa}
\rho_{\kappa\nu} C_{\nu\kappa}|
\label{u48}
\\
W & := & |\sum_{\kappa\lambda=1}^D 
\rho^2_{\kappa\lambda} C^2_{\lambda\kappa}|
\label{u49}
\end{eqnarray}
Similarly as in (\ref{u9b}), the sums over $\lambda$ and
$\nu$ in (\ref{u48}) can be readily performed,
yielding
\begin{eqnarray}
V & = & |\sum_{\kappa=1}^D 
(\langle\kappa|\rho C|\kappa\rangle)^2|
\leq \sum_{\kappa=1}^D
|\langle\kappa|\rho C|\kappa\rangle|^2
\ .
\label{u50}
\end{eqnarray}
Choosing $|\psi\rangle=|\kappa\rangle$ 
and $|\phi\rangle=C|\kappa\rangle$ in (\ref{u4m}),
it follows that
\begin{eqnarray}
V & \leq & 
\sum_{\kappa=1}^D
\langle\kappa|\rho|\kappa\rangle
\langle\kappa|C\rho C|\kappa\rangle
\nonumber
\\
& \leq &
\max_n \rho_{nn}(0)\  \tr\{C\rho C\}
\ .
\label{u51}
\end{eqnarray}
Evaluating the trace by means of the eigenbasis of $C$,
one sees that $\tr\{C\rho C\}\leq\norm{C}^2$.
With (\ref{u4f}) we thus can conclude that
\begin{eqnarray}
V & \leq & 
4\, \norm{A}^2\, \max_n \rho_{nn}(0)
\ .
\label{u52}
\end{eqnarray}
To evaluate (\ref{u49}), one proceeds similarly
as in (\ref{u34})-(\ref{u41}), yielding
\begin{eqnarray}
W \leq \norm{C}^2\tr\rhobar^2
\leq 4\,\norm{A}^2\tr\rhobar^2
\label{u53}
\end{eqnarray}
In combination with (\ref{u47}) and (\ref{u52}), 
we thus can conclude that
\begin{eqnarray}
|S_{11}| & \leq & 
4\, \norm{A}^2\, [\tr\rhobar^2 + \max_n \rho_{nn}(0)]
\ .
\label{u54}
\end{eqnarray}

Along the same line of reasoning,
one finds for $a=13$ that
\begin{eqnarray}
|S_{13}| & \leq & 
4\, \norm{A}^2\, [\tr\rhobar^2 + \max_n \rho_{nn}(0)]
\ .
\label{u55}
\end{eqnarray}

Taking into account (\ref{u19}) and (\ref{u20})
we finally can infer from (\ref{u41}), (\ref{u44}),
(\ref{u54}), (\ref{u55}) that
\begin{eqnarray}
\sum_{a=9}^{14} |S_a| \leq 
\norm{A}^2\, [18\, \tr\rhobar^2 + 16 \,\max_n \rho_{nn}(0)]
\ .
\label{u56}
\end{eqnarray}

%%%%%%%%%%%%%%%%%%%%%%%%%%%%%%%%%%%%
\subsection{Evaluation of $S_{15}$}
We start with the trivial identities
\begin{eqnarray}
S_{15} & = & X+Y+Z+R^2
\label{u60}
\\
X & := & S_{15}-\alpha
\label{u61}
\\
Y & := & \alpha-\beta
\label{u62}
\\
Z & := & \beta-R^2
\label{u63}
\\
\alpha & := & 
(\tr\{\rho(0)C\})^2
\frac{(D-4)!}{D!}\,
{\sum_{klmn\!\!}}'e^{ln}_{km}
\label{u64}
\\
\beta & := & 
(\tr\{\rho(0)C\})^2
\frac{(D-4)!}{D!}\,
{\sum_{kl}}'{\sum_{mn}}'e^{ln}_{km}
\ .
\label{u65}
\end{eqnarray}

For $a=15$, all four indices in (\ref{t2}) are
pairwise distinct, and with (\ref{u15})
it follows that
\begin{eqnarray}
S_{15} 
& = &  
\!\! {\sum_{klmn\!\!}}'
[h(\pi(k),\pi(l),\pi(m),\pi(n))]_\Pi \ e^{ln}_{km}
\nonumber
\\
& = & 
\!\! {\sum_{klmn\!\!}}' e^{ln}_{km}
\frac{(D-4)!}{D!}\,
{\sum_{\kappa\lambda\mu\nu\!\!}}' 
h(\kappa,\lambda,\mu,\nu)
\ .
\label{u66}
\end{eqnarray}
In the last step we exploited (\ref{s7}) with $K=4$.

Upon comparison of (\ref{u9}) and (\ref{u9i})
one sees that $\tr\{\rho(0)C\}=
\sum_{\mu\nu\!\!}'\,
g(\mu,\nu)$
and hence (\ref{u64}) 
can be rewritten with (\ref{u11}) as
\begin{eqnarray}
\alpha & = & 
\!\! {\sum_{klmn\!\!}}' e^{ln}_{km}
\frac{(D-4)!}{D!}\,
{\sum_{\kappa\lambda}}' {\sum_{\mu\nu}}'
h(\kappa,\lambda,\mu,\nu)
\label{u67}
\end{eqnarray}
Introducing (\ref{u66}) and (\ref{u67}) into (\ref{u61})
yields
\begin{eqnarray}
X & = & X_1\, X_2
\label{u68}
\\
X_1 & := & 
{\sum_{klmn\!\!}}' e^{ln}_{km}
\frac{(D-4)!}{D!}
\label{u69}
\\
X_2 & := & \!\!\!\!\! 
\sum_{klmn\in I_{15}} \!\!\!\!
h(k,l,m,n)
-
\!\!\!\! \sum_{klmn\in J}\!\!\!\!  h(k,l,m,n)
\ .
\label{u70}
\end{eqnarray}
For later convenience,
we employed $k,l,m,n$ instead
of $\kappa,\lambda,\mu,\nu$ as summation indices
in (\ref{u70}).
Furthermore, the sets $I_{15}$ and $J$ 
are defined in (\ref{t2}) and (\ref{t3}),
respectively.

With (\ref{u13}) and (\ref{u4}) it follows that
$|e^{ln}_{km}|=1$.
Since the primed sum over $k,l,m,n$ in
(\ref{u69})
consists of $D!/(D-4)!$ summands, 
we can conclude that
\begin{eqnarray}
|X_1| \leq 1
\ .
\label{u71}
\end{eqnarray}

As observed below (\ref{t3}), the set $J$ is the
disjoint union of $I_{9}$,...,$I_{15}$, implying
that (\ref{u70}) can be rewritten as
\begin{eqnarray}
X_2 & = & -\, \sum_{a=9}^{14} S'_{a}
\label{u72}
\\
S'_a & := & \sum_{klmn\in I_{a\, }} \!\!
h(k,l,m,n)
\ .
\label{u73}
\end{eqnarray}
With (\ref{u13}) and (\ref{u4}) it follows that
$e^{ln}_{km}=1$ for $t=0$.
Hence, $S'_a$ from (\ref{u73}) can be identified
with $S_a$ from (\ref{u15}) 
by taking into account (\ref{s7})
and setting $t=0$.
Since the estimate from (\ref{u56}) is valid
independently of $t$, we can conclude 
that
\begin{eqnarray}
\sum_{a=9}^{14} |S'_a| \leq 
\norm{A}^2\, [18\, \tr\rhobar^2 + 16 \,\max_n \rho_{nn}(0)]
\ .
\label{u74}
\end{eqnarray}
Altogether, (\ref{u68}) with
(\ref{u71})-(\ref{u74}) implies
\begin{eqnarray}
X \leq \norm{A}^2\, [18\, \tr\rhobar^2 + 16 \,\max_n \rho_{nn}(0)]
\ .
\label{u75}
\end{eqnarray}

In the same vein, we rewrite (\ref{u62}) 
with (\ref{u64}) and (\ref{u65}) as
\begin{eqnarray}
Y & = & Y_1\, Y_2
\label{u76}
\\
Y_1 & := &
(\tr\{\rho(0)C\})^2 \frac{(D-4)!}{D!}
\label{u77}
\\
Y_2 & := &
\sum_{klmn\in I_{15}} \!\!\!
e^{ln}_{km}
-
\!\!\!\! \sum_{klmn\in J}\!\!\!  e^{ln}_{km}
\ ,
\label{u78}
\end{eqnarray}
where $I_{15}$ and $J$ are defined in
(\ref{t2}) and (\ref{t3}), respectively.

Evaluating the trace by means of the eigenbasis of $C$,
one sees that $|\tr\{\rho(0)C\}| \leq \norm{C}$.
With (\ref{u4f}) we thus can conclude that
\begin{eqnarray}
|Y_1| & \leq &
4\,\norm{A}^2 \frac{(D-4)!}{D!}
\ .
\label{u79}
\end{eqnarray}

Since $|e^{ln}_{km}|=1$ (see (\ref{u13}) and (\ref{u4}))
and since $J$ is the disjoint union of $I_{9}$,...,$I_{15}$
(see below (\ref{t3})) we can conclude from (\ref{u78}) that
\begin{eqnarray}
|Y_2| & \leq & \sum_{a=9}^{14} |I_a|
\ ,
\label{u80}
\end{eqnarray}
where $|I_a|$ denotes the size 
(number of elements) of the set $I_a$.
From (\ref{t2}) one readily concludes that
$|I_a|=D(D-1)$ for $a=9,10$ and
$|I_a|=D(D-1)(D-2)$ for $a=11,...,14$,
yielding
\begin{eqnarray}
\sum_{a=9}^{14} |I_a| = 4D(D-1)(D-3/2)
\ .
\label{u81}
\end{eqnarray}

Introducing (\ref{u79})-(\ref{u81}) into (\ref{u76})
implies
\begin{eqnarray}
|Y| & = & \norm{A}^2 q(D)/D
\label{u82}
\\
q(D) & := & 16 \frac{D}{D-2}\frac{D-3/2}{D-3} 
\ .
\label{u83}
\end{eqnarray}
Observing that the last two factors on the 
right hand side of (\ref{u83})
are both decreasing functions of $D$, we obtain
$q(D)\leq q(6)=36$ for $D\geq 6$.
% or,since $D$ is integer, $D>5$.
Altogether, we thus arrive at
\begin{eqnarray}
|Y| & \leq & 36 \,\norm{A}^2/D\ \ \mbox{for}\ D\geq 6
\ .
\label{u84}
\end{eqnarray}

With the help of (\ref{u9i}) we can conclude 
from (\ref{u63}) and (\ref{u65}) that
\begin{eqnarray}
Z & = & Z_1\, Z_2
\label{u85}
\\
Z_1 & := & (\tr\{\rho(0)C\})^2
{\sum_{kl}}'{\sum_{mn}}'e^{ln}_{km}
\label{u86}
\\
Z_2 & := & \frac{(D-4)!}{D!}-
\frac{1}{D^2(D-1)^2}
\label{u87}
\end{eqnarray}
As before, we exploit that $|\tr\{\rho(0)C\}| \leq 2\norm{A}$,
that the two double sums in (\ref{u86}) give rise to
$D^2(D-1)^2$ summands, and that $|e^{ln}_{km}|=1$
to infer
\begin{eqnarray}
|Z_1| & \leq & 4\,\norm{A}^2 D^2(D-1)^2
\ .
\label{u88}
\end{eqnarray}
Introducing (\ref{u87}) and (\ref{u88}) into
(\ref{u85}) readily yields the same upper
bound for $|Z|$ as for $|Y|$ in (\ref{u82}).
Like in (\ref{u84}), we thus obtain
\begin{eqnarray}
|Z| & \leq & 36 \,\norm{A}^2/D\ \ \mbox{for}\ D\geq 6
\ .
\label{u89}
\end{eqnarray}
%%%%%%%%%%%%%%%%%%%%%%%%%%%%%%%%%%%%
\subsection{Final result}
From the definition (\ref{u4a}) of $\xi_\pi(t)$ we
can infer that
\begin{eqnarray}
[\xi^2_\pi(t)]_\Pi & = &
\left[(\langle A\rangle_{\!\rho(t)})^2\right]_\Pi
-
\left([\langle A\rangle_{\!\rho(t)}]_\Pi\right)^2
\label{u90}
\end{eqnarray}
and from (\ref{u5}) that
\begin{eqnarray}
\left([\langle A\rangle_{\!\rho(t)}]_\Pi\right)^2
= Q^2+2QR+R^2
\ .
\label{u91}
\end{eqnarray}
Introducing (\ref{u32}) and (\ref{u60}) into
(\ref{u14}) yields
\begin{eqnarray}
\left[(\langle A\rangle_{\!\rho(t)})^2\right]_\Pi
= & & Q^2+2QR+\sum_{a=9}^{14}S_a +
\nonumber
\\
& & +X+Y+Z+R^2
 \ .
\label{u92}
\end{eqnarray}
From (\ref{u90})-(\ref{u92}) we can conclude that
\begin{eqnarray}
[\xi^2_\pi(t)]_\Pi & = & \sum_{a=9}^{14}S_a+X+Y+Z
\label{u93}
\end{eqnarray}
and with (\ref{u56}), (\ref{u75}), (\ref{u84}), (\ref{u89})
that
\begin{eqnarray}
[\xi^2_\pi(t)]_\Pi \leq 
4 \norm{A}^2\, [9\tr\rhobar^2 + 8\max_n \,\rho_{nn}(0)+18/D]
\label{u94}
\end{eqnarray}
for $D\geq 6$.
Observing that $D^{-1}\leq \tr\rhobar^2$ and
(\ref{u4g}), we obtain
\begin{eqnarray}
[\xi^2_\pi(t)]_\Pi & \leq & 
\da^2\, [27\, \tr\rhobar^2 + 8 \,\max_n \rho_{nn}(0)]
\label{u95}
\end{eqnarray}
for $D\geq 6$.
Finally, we exploit the inequality
\begin{eqnarray}
\tr\rhobar^2 = \sum_n \rho^2_{nn}(0)\leq \max_n \rho_{nn}(0)
\label{u96}
\end{eqnarray}
and $35<6^2$ to rewrite (\ref{u95}) as
\begin{eqnarray}
[\xi^2_\pi(t)]_\Pi & \leq & 
(6\da)^2 \max_n \rho_{nn}(0)
\label{u97}
\end{eqnarray}
for $D\geq 6$.
This is identical to the second relation in
(6) of the main text.

We also tried to derive a slightly stronger 
version of (\ref{u97}) with $\tr\rhobar^2$ 
instead of $\max_n \rho_{nn}(0)$ appearing 
on the right hand side.
The main problem is the last summand 
in (\ref{u95}), whose origin is the estimate
in (\ref{u50}), (\ref{u51}).
Unfortunately, we did not succeed to
find an analogous estimate but in 
terms of $\tr\rhobar^2$ instead of 
$\max_n \rho_{nn}(0)$.

\end{document}